\documentclass[aps,showpacs,prx,amssymb,amsmath,twocolumn]{revtex4-2}
\usepackage{graphicx}
\usepackage{amssymb}
\usepackage{amsmath}
\usepackage{amsthm}
\usepackage{bm}
\usepackage{xcolor} 
\usepackage{array}
\usepackage{multirow}
\usepackage{tabularx}
\usepackage{booktabs}
\usepackage{textcomp}
\usepackage{mathtools}
\usepackage{hyperref}
\usepackage{qcircuit}
\usepackage{bbm}
\usepackage{cancel}
\usepackage{comment}
\usepackage{physics}

\begin{document}
\title{Rydberg-atom graphs for quadratic unconstrained binary optimization problems}
\author{Andrew Byun$^1$, Junwoo Jung$^1$, Kangheun Kim$^1$, Minhyuk Kim$^{1,2}$, Seokho Jeong$^1$, Heejeong Jeong$^3$ and Jaewook Ahn$^1$}
\address{$^1$Department of Physics, Korea Advanced Institute of Science and Technology (KAIST), Daejeon 34141, Korea}
\address{$^2$Department of Physics, Korea University, Seoul 02841, Korea}
\address{$^3$Qunova Computing Inc., Daejeon 34130, Korea}

\begin{abstract} \noindent
There is a growing interest in harnessing the potential of the Rydberg-atom system to address complex combinatorial optimization challenges. Here we present an experimental demonstration of how the quadratic unconstrained binary optimization (QUBO) problem can be effectively addressed using Rydberg-atom graphs. The Rydberg-atom graphs are configurations of neutral atoms organized into mathematical graphs, facilitated by programmable optical tweezers, and designed to exhibit many-body ground states that correspond to the maximum independent set (MIS) of their respective graphs. We have developed four elementary Rydberg-atom subgraph components, not only to eliminate the need of local control but also to be robust against interatomic distance errors, while serving as the building blocks sufficient for formulating generic QUBO graphs. To validate the feasibility of our approach, we have conducted a series of Rydberg-atom experiments selected to demonstrate proof-of-concept operations of these building blocks. These experiments illustrate how these components can be used to programmatically encode the QUBO problems to Rydberg-atom graphs and, by measuring their many-body ground states, how their QUBO solutions are determined subsequently.
\end{abstract}

\maketitle

\section{Introduction}
Quantum computing~\cite{Nielsen2010} has sparked considerable interest in recent years, because of its potential to solve combinatorial optimization problems, which demand significant classical resources~\cite{Papadimitriou1998}, by leveraging quantum resources for vastly greater information capacity than classical counterparts~\cite{Albash2018}. Combinatorial optimization problems are often cast as graph problems~\cite{Papadimitriou1998}, making Ising-spin graphs as a prototypical platform for solving NP-problems~\cite{Lucas2014_Ising} and Rydberg-atom system~\cite{Saffman2010_review, Browaeys2020_review} is among the controllable Ising-type quantum simulators~\cite{Bernien2017}. Arrangements of Rydberg atoms using optical tweezers have enabled versatile construction of arbitrary graphs~\cite{Pichler2018_MIS, Kim2020_3Dgraph}. Furthermore, Rydberg quantum wires, additional atomic chains, offer a solution to circumvent geometric and topological constraints in physical graph representation~\cite{Qiu2020_wire, Kim2022_wire, Byun2022, Jeong2023_3SAT}, facilitating the establishment of both planar and non-planar graphs within the Rydberg atom system. Notably, Rydberg atom graph have recently demonstrated its capacity as a graph problem solver, successfully addressing the maximum independent set (MIS) problem~\cite{Kim2022_wire, Byun2022, Ebadi2022_MIS}, the MaxCut problem~\cite{Graham2022_MAXCUT}, and the 3-satisfiability (SAT) problem~\cite{Jeong2023_3SAT}.

Quadratic unconstrained binary optimization (QUBO) problem aims to find the optimum $N$-bit binary string (bitstring) $\mathbf{x}=(x_1,x_2,\cdots,x_N)$, where each $x_i \in \{0,1\}$, that minimizes a quadratic cost function $f(\mathbf{x})$, given by
\begin{eqnarray}\label{QUBO_costfunction}
f(\mathbf{x})&=&\sum_{i=1}^N Q_{ii}x_i+\sum_{i<j} Q_{ij} x_i x_j,
\end{eqnarray}
where ${Q}$ is a real-valued $N\times N$ square matrix. The elements of ${Q}_{ii}$ and $Q_{ij}$ respectively define linear and quadratic terms in the cost function. QUBO is gaining significant attention due to its simple and versatile representation~\cite{Glover2018_qubo}. Solving the optimization problem in the QUBO form is mostly attempted with D-wave machine~\cite{dwave2017, dwave2017_2, dwave2018_2, dwave2019, dwave2020, dwave2021_2, dwave2021_3, dwave2022, dwave2022_2, dwave2023}. In the context of Rydberg atom systems, there are proposals for QUBO programming~\cite{Keating2013_rydbergqubo, Nguyen2022_mwis, Lanthaler2023_mwis}; however, technical challenges arise from the necessity of full local control over individual qubits~\cite{Graham2022_MAXCUT, Labuhn2014_addressing, Omran2019_20addressing} and the distance-sensitive nature of atom-atom interactions~\cite{Beguin2013, Ravets2014, Jo2020, Chew2022}.

In this work, we consider a global-control version of Rydberg-atom implementation of the QUBO problem. We present a set of Rydberg-atom experiments to implement the QUBO problem using Rydberg atom graphs. We extend the concept of the many-body ground-state equivalence of the Rydberg quantum wire~\cite{Qiu2020_wire, Kim2022_wire} to express general QUBO cost functions in integer forms. Since all real number coefficients can be effectively approximated as rational numbers and multiplying the cost function coefficients by a positive scaling factor does not affect the QUBO solution, all coefficients are in principle representable as integers. We develop four Rydberg-atom QUBO gadgets, which are sufficient for the QUBO implementation of a general cost function. It shall be shown that all many-body ground state configurations stemming from the Rydberg quantum wire under a global laser field can be mapped to QUBO solutions, eliminating the need for local control techniques and distance-dependent interactions. 

The remaining of the paper is structured as follows: In Section.~\ref{GRAPH}, we introduce our four elementary and complete building blocks: the data qubit, offset qubit, and even- and odd-number quantum wires. We illustrate how each building block represents distinct Hamiltonian terms. By employing these building blocks, we can effectively express QUBO cost functions in integer formats. This implies that Rydberg-atom graphs possess the capability to both implement QUBO cost functions and express weighted graphs. The experimental procedure is detailed in Section.~\ref{Experiment}, and two- and three-variable QUBO problem examples are presented in Section.~\ref{Result}. Lastly, we present our conclusions in Section.~\ref{Discussions}.

\begin{figure*}[t]
    \centering
    \includegraphics[width=2.0\columnwidth]{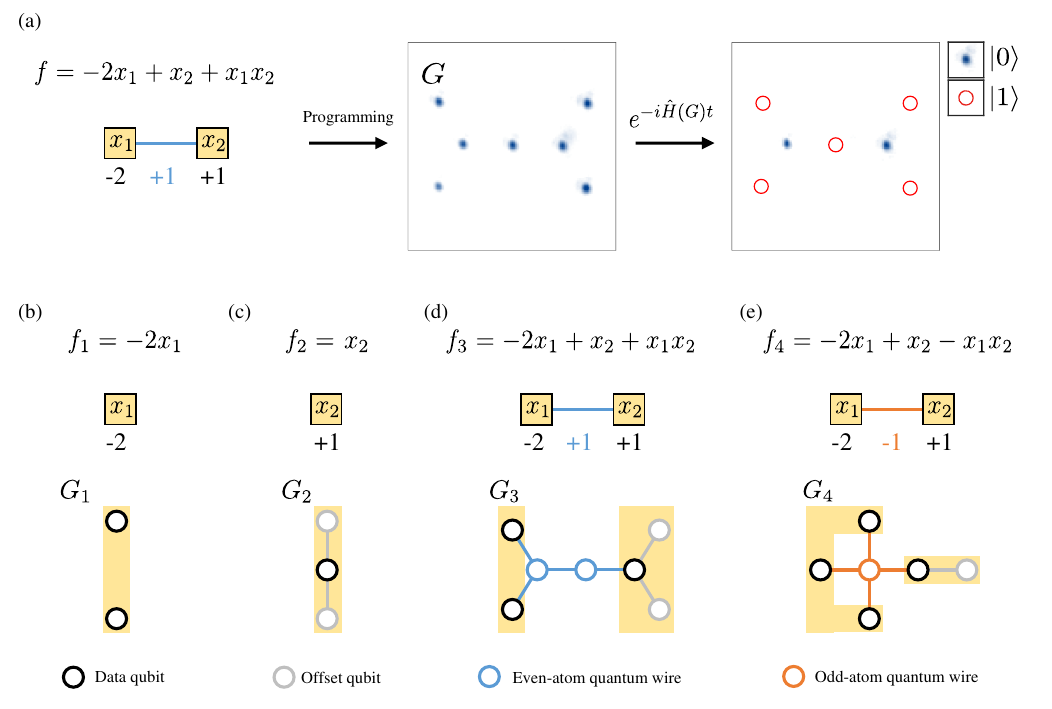}
    \caption{(a) A procedure of the Rydberg-atom graph implementation of the QUBO problem: For a given QUBO cost function $f$, a corresponding weighted graph is programmed with a Rydberg-atom graph $G$ and subsequently the QUBO solution of $f$ is obtained from the many-body ground state of $G$. (b-e) Rydberg-atom graph examples for four elementary building blocks: (b) The data-qubit subgraph $G_1$ for the QUBO cost function $f_1=-2x_1$ with a negative linear term. (c) The data-and-offset subgraph $G_2$ for $f_2=x_2$ with a positive linear term. (d) The ``even-atom'' quantum wire in $G_3$ for $f_3=f_1+f_2+x_1x_2=-2x_1+x_2+x_1x_2$ with a positive quadratic term. (e) The ``odd-atom'' quantum wire in $G_4$ for  $f_4=f_1+f_2-x_1x_2=-2x_1+x_2-x_1x_2$ with a negative quadratic term.
}
    \label{Fig1}
\end{figure*}

\section{Rydberg-atom graph for QUBO} \label{GRAPH} \noindent
A Rydberg-atom graph is a group of atoms spatially arranged to represent a mathematical graph $G(V,E)$, in which atoms and Rydberg-blockaded inter-atom interactions correspond to the vertices ($V$) and edges ($E$), respectively~\cite{Pichler2018_MIS, Kim2020_3Dgraph}. The Hamiltonian is given by (in the unit of $\hbar=1$)
\begin{equation}\label{HRyd}
\hat{H}_G=\frac{\Omega}{2} \sum _{i\in V} \hat{\sigma}_{x,i}-\Delta \sum _{i\in V} \hat{n}_{i}+\sum _{(i,j)\in E} U_{i,j} \hat{n}_{i} \hat{n}_{j},
\end{equation}
where $U_{i,j}=C_6/|\mathbf{r}_i-\mathbf{r}_j|^6$ is the van der Waals interaction with coefficient $C_6$ between a Rydberg-blockaded atom pair $i,j$, i.e., $|\mathbf{r}_{i}-\mathbf{r}_{j}|<(C_6/\sqrt{\Omega^2+\Delta^2})^{1/6}=d_R$~\cite{Jaksch2000_blockade, UrbanNatPhys2009_blockade, GaetanNatPhys2009_blockade}. $\Omega$ and $\Delta$ are the control parameters of the Rydberg-excitation laser, to be detailed in Section.~\ref{Experiment}. 
$\hat{\sigma}_x=\left|0\right>\left<1\right|+\left|1\right>\left<0\right|$ and $\hat{n}=\left|1\right>\left<1\right|$ are bit-flip and number operators defined for the ground and Rydberg states respectively denoted by $\left|0\right>$ and $\left|1\right>$. Under the ``MIS condition'', given by $0<\Delta < U$ and $\Omega \rightarrow 0$, the second term of Equation.~\ref{HRyd}, which favors maximum Rydberg atoms, and the third term, which restricts Rydberg-atom pairs, compete with each other, so the MIS problem, a special kind of graph-coloring problem that involves finding the largest set of non-adjacent vertices in a graph, is naturally solved by the Rydberg-atom graph~\cite{Pichler2018_MIS}.

We aim to program the QUBO cost function in Eq.~\ref{QUBO_costfunction} using a Rydberg-atom graph so that QUBO solutions can be obtained at the minimum of the following Hamiltonian:
\begin{eqnarray}\label{RydbergQUBO}
\hat{H}_{G} \stackrel{g}{=} \hat{H}_{\rm QUBO}= \sum_{i=1}^N Q_{ii} \hat{n}_i+\sum_{i<j} Q_{ij} \hat{n}_i \hat{n}_j,
\end{eqnarray}
where $Q$ is an integer-valued $N \times N$ matrix and the symbol $\stackrel{g}{=}$ represents the many-body ground-state equivalence between the left-hand side and right-hand side terms, of which the concept is developed in Rydberg quantum wires~\cite{Qiu2020_wire, Kim2022_wire}, meaning that the left and right Hamiltonians share the same many-body ground states (not the same energy). 

In Figure.~\ref{Fig1}(a), the procedure to obtain the solution of a test QUBO problem  $f=-2x_1+x_2+x_1x_2$ is outlined as an example. The equivalent QUBO ``weighted'' graph is shown below the function in the figure, in which the vertex weights for vertices, $-2x_1$ and $x_2$, are $-2$ and $+1$, respectively, and the edge weight for the edge $x_1 x_2$ is $+1$. As to be explained below, the first term, $f_1=-2x_1$, is programmable with a data-qubit graph, shown in Figure.~\ref{Fig1}(b), the second term,  $f_2=x_2$, with an offset-qubit graph in Figure.~\ref{Fig1}(c), and the third term, $x_1x_2$, with a quantum wire in Figure.~\ref{Fig1}(d). So, the function $f=f_1+f_2+x_1x_2$ can be programmed with the Rydberg-atom graph $G$, shown in the middle panel of Figure.~\ref{Fig1}(a) and the many-body ground state of $\hat H(G)$ is as in the right panel, resulting that $f(x_1=1,x_2)=-2$ is the cost function minimum. 

There are four elementary building blocks that are sufficient and necessary for Rydberg-atom implementation of a general QUBO function. They are the data-qubit subgraph (for a negative vertex weight), the offset-qubit subgraph (for a positive vertex weight), and ``even-atoms'' and ``odd-atoms'' Rydberg quantum wires (for edge weights). Now we construct each of them with a Rydberg-atom subgraph and quantum wire. 

{\it (1) Data-qubit subgraph}: A negative-coefficient linear term, e.g., $Q_{11}=-2$ in $f_1=-2x_1$ in Figure.~\ref{Fig1}(b), of a QUBO cost function can be programmed with a data-qubit subgraph, which consists of multiple ($|Q_{11}|=2$) atoms that are not directly coupled with each other, as shown with thick circles in Figure.~\ref{Fig1}(b), because these atoms contribute to the Hamiltonian as
\begin{equation}\label{Hdata}
\frac{\hat{H}^{\rm data}_{i} }{\Delta}= - |Q_{ii}| \hat{n}_{i}=Q_{ii} \hat{n}_{i},
\end{equation}
where $|Q_{ii}|$ is the number of atoms for variable $x_i$, favoring $x_i=1$ for Hamiltonian minimum.

{\it (2) Data-and-offset subgraph}: A positive-coefficient linear term, e.g., $Q_{22}=+1$ in $f_2=x_2$ in Figure.~\ref{Fig1}(c), is programmable with an data-and-offset subgraph, consisting of multiple ($Q_{22} +1=2$ in this case) 
auxiliary atoms coupling with the atom representing the variable $x_2$, as shown in Figure.~\ref{Fig1}(c), where the gray circles are the auxiliary atoms. When the $i$-th data qubit is coupled with a $Q_{ii}+1$ number of auxiliary atoms, this data-and-offset subgraph effectively increases the Hamiltonian by
\begin{equation}\label{Hoffset}
\frac{\hat{H}^{\rm offset}_{i} }{\Delta} = -\hat{n}_{i} + (Q_{ii}+1)\hat{n}_{i} = Q_{ii} \hat{n}_{i},
\end{equation}

The quadratic terms of a QUBO cost function can be programmed with Rydberg quantum wires~\cite{Qiu2020_wire,Kim2022_wire, Byun2022, Jeong2023_3SAT}.  A Rydberg quantum wire is a chain of auxiliary atoms to couple a pair of remotely located data-qubit atoms. An ``even-atom'' quantum wire, which uses an even number of auxillary atoms, ensures the many-body ground-state equivalence between the quantum-wired data-qubit atoms and directly-coupled data-qubit atoms~\cite{Kim2022_wire}. Table~\ref{Table1} shows the ground-state energies of the coupled system of two data-qubit atoms and an ``even-atom'' quantum wire, showing the equivalence, given by
\begin{equation} \label{EvenWire}
    \hat{H}_{ij}^{\rm even} \stackrel{g}{=} \Delta \hat{n}_i \hat{n}_j.
\end{equation}
On the other hand, when an odd number of auxiliary atoms are used for an ``odd-atom'' Rydberg quantum wire, the ground-state energies of the coupled system are given as in Table~\ref{Table2}. So the Hamiltonian of the system of the two data-qubit atoms and an ``odd-atom'' quantum can be summarized as
\begin{equation} \label{OddWire}
    \hat{H}^{\rm odd} \stackrel{g}{=} \Delta (\hat{n}_i + \hat{n}_j -  \hat{n}_i\hat{n}_j).
\end{equation}

\begin{table}[bt]
\caption{Ground-state energies of two data-qubit atoms ($x_i,x_j$) coupled with an ``even-atom'' quantum wire of an even number $2M$ of auxiliary atoms}
\begin{ruledtabular}
\centering
\begin{tabular}{clr}
$\left|x_i x_j\right>$ & Gound-state configurations & Energy\\ 
\hline
$\left|11\right>$ & $ \left|1\right>\otimes \left|01\cdots 00 \cdots 10\right> \otimes \left|1\right>$ & $(1-M) \Delta$ \\
$\left|10\right>$ & $ \left|1\right>\otimes \left|01\right>^{\otimes M} \otimes \left|0\right>$ & $-M \Delta $ \\
$\left|01\right>$ & $ \left|0\right>\otimes \left|10\right>^{\otimes M} \otimes \left|1\right> $ & $-M\Delta$ \\
$\left|00\right>$ & $ \left|0\right>\otimes \left|10\right>^{\otimes M} \otimes \left|0\right>$ or $\left|0\right> \otimes \left|10\cdots01\right>\otimes \left|0\right>$ & $-M \Delta$ \\
\end{tabular}
\end{ruledtabular}
\label{Table1}
\end{table}

\begin{table}[bt]
\caption{Ground-state energies of two data-qubit atoms ($x_i,x_j$) coupled with an ``odd-atom'' quantum wire of an odd number $2M+1$ of auxiliary atoms}
\begin{ruledtabular}
\centering
\begin{tabular}{clr}
$\left|x_ix_j\right>$ & Ground-state configurations & Energy\\ 
\hline
$\left|11\right>$ & $ \left|1\right> \otimes  \left|01\cdots 0 \cdots 10\right> \otimes \left|1\right> $ & $-M\Delta $  \\
$\left|10\right>$ & $ \left|1\right> \otimes \left|01\cdots 0 \cdots 10\right> \otimes \left|0\right> $ & $-M\Delta$ \\
$\left|01\right>$ & $ \left|0\right> \otimes \left|01\cdots 0 \cdots 10\right> \otimes \left|1\right> $ & $-M\Delta$ \\
$\left|00\right>$ & $ \left|0\right> \otimes (\left|10\right>^{\otimes M} \otimes\left|1\right> )\otimes \left|0\right>$ & $-(M+1)\Delta$ \\
\end{tabular}
\end{ruledtabular}
\label{Table2}
\end{table}

{\it (3) ``Even-atom'' quantum wire}: A positive-coefficient quadratic term, e.g., $Q_{12}=+1$ in $f_3=f_1+f_2+x_1x_2$ in Figure.~\ref{Fig1}(d), of a QUBO cost function can be programmed with a number ($Q_{12}=1$ in this case) of ``even-atom'' Rydberg quantum wires, by using the equivalence in Eq.~\ref{EvenWire}, as shown in Fig.~\ref{Fig1}(d), where the blue circles denote the quantum wire. When a $Q_{ij}$-number of quantum wires couple two data qubits, $i,j$, the Hamiltonian is given by
\begin{equation}\label{Heven}
\frac{\hat{H}^{\rm even}_{ij} }{ \Delta} \stackrel{g}{=}  Q_{ij} \hat{n}_i\hat{n}_j.
\end{equation}

{\it (4) ``Odd-atom'' quantum wire}: A negative-coefficient quadratic term, e.g., $Q_{12}=-1$ in $f_4=(f_1-x_1)+(f_2-x_2)+(x_1+x_2-x_1x_2)$ in Figure.~\ref{Fig1}(e), requires not only ``odd-atom'' quantum wires but also data(offset)-qubit graphs, where the latter eliminate the additional linear terms occurring in Eq.~\ref{OddWire}. So, for a $Q_{ij}=-|Q_{ij}|$ quadratic term, we need $|Q_{ij}|$-number of ``odd-atom'' quantum wires and data-qubit graphs for $x_i$ and $x_j$ variables. In Figure.~\ref{Fig1}(e), to express $f_1-x_1$ and $f_2-x_2$, one more data qubit is added and one offset qubit is deleted. The resulting Hamiltonian is given by
\begin{equation}\label{Hodd}
\frac{\hat{H}^{\rm odd + data}_{ij} }{ \Delta} \stackrel{g}{=}  -|Q_{ij}| \hat{n}_i\hat{n}_j= Q_{ij} \hat{n}_i\hat{n}_j.
\end{equation}

As a result, by combining the four building blocks, we can program general QUBO Hamiltonians as
\begin{equation}
\hat{H}_{\rm QUBO} \stackrel{g}{=} \sum_{i=1}^N \hat{H}^{\rm qubit}_{i} + \sum_{i<j} \hat{H}^{\rm wire}_{(i,j)},
\end{equation}
where $\hat{H}^{\rm qubit}_{i}$ is the qubit-class Hamiltonians for the data-qubit and data-and-offset subgraphs in Eqs.~\ref{Hdata} and \ref{Hoffset}, respectively, and $\hat{H}^{\rm wire}_{(i,j)}$ is the ``even-atom'' and ``odd-atom'' quantum-wire Hamiltonians in Eqs.~\ref{Heven} and \ref{Hodd}, respectively.

\section{Experimental procedure} \label{Experiment} \noindent
We now describe how to implement those Rydgerg-atom graphs experimentally and then how to obtain their many-body ground states (to solve the programmed QUBO problems).

Rubidium atoms ($^{87}$Rb) are laser-cooled down to 35~$\mu$K in a magneto-optical trap and individually trapped by optical tweezers controlled by a spatial light modulator (Meadowlark Optics ODPDM512)~\cite{Nogrette2014_tweezer, Kim2016_rearrangement, Kim2019_zeropadding}. For deterministic loading of all atoms on spatial positions for Rydberg-atom graphs, acousto-optic deflectors (AA Opto Electronic DTSXY-400-800) are used for dynamic control of the optical tweezers~\cite{Barredo2016_rearr, Endres2016_rearragnge}. Each of the atoms is initially optically pumped to the ground state $\left|0\right>=\left|5S_{1/2}, F=2, m_F=2\right>$, which is then optically coupled to the Rydberg state $\left|1\right>=\left|71S_{1/2}, J=1/2, m_J=1/2\right>$, by the two-photon absorption scheme through the intermediate state $\left|m\right>=\left|5P_{3/2}, F'=3, m_{F'}=3\right>$.

The coupling strengths of the two-photon transitions are $\Omega_{0m}=106$~$(2\pi)$MHz and $\Omega_{m1}=12$~$(2\pi)$MHz, respectively, and the intermediate detuning is $\Delta_{m}=670$~$(2\pi)$MHz. So, the resulting Rabi frequency is $\Omega_0=\Omega_{0m}\Omega_{m1}/2\Delta_m=0.96$~$(2\pi)$MHz. The Rydberg-excitation lasers are a homemade 780-nm external-cavity diode laser ($\sigma^+$) and a Toptica TA-SHG Pro 480-nm laser ($\sigma^-$), from which the beams counter-propagate with each other and PDH-locked to 
{an ultra-low-expansion} Fabry-Perot cavity (Stable Laser Systems, ATF-6010-4) of finesse $\mathcal{F}=15,000$. The laser phase noise are peaked at 
{$0.34$~MHz for $3\times 10^{-7}$~Hz$^{-1}$ of 780-nm laser and at $0.22$~MHz for $5\times 10^{-8}$~Hz$^{-1}$ in 480-nm laser. }
With a constructed Rydberg-atom graph for $G(V,E)$, we change the atom states from their initial $\left|0\right>^{\otimes N}$ state to the many-body ground state of the Hamiltonian $\hat H_G(\Omega=0, \Delta=\Delta_f)$ in Eq.~\ref{HRyd}. This is achieved by progressively altering the Hamiltonian $\hat H_G(\Omega, \Delta)$ from $\hat H_G(\Omega=0, \Delta=\Delta_i)$ to $\hat H_G(\Omega_0, \Delta=\Delta_i)$, and then subsequently to 
$\hat H_G(\Omega_0, \Delta=\Delta_f)$, and finally to $\hat H_G(\Omega=0, \Delta=\Delta_f)$. In more detail, in the first step, from $t=0$ to $t=0.1 T$, the detuing remains constant at $\Delta(t=0-0.1T)=\Delta_i$ and the Rabi frequency is linearly changed from $\Omega(t=0)=0$ to $\Omega(t=0.1T)=\Omega_0$. In the second step, from $t=0.1T$ to $t=0.9T$, the Rabi frequency remains constant at $\Omega(t=0.1T-0.9T)=\Omega_0$ and the detuning is linearly changed from 
$\Delta(t=0.1T)=\Delta_i$ to $\Delta(t=0.9T)=\Delta_f$. In the final step, from $t=0.9T$ to $t=T$, the detuing remains constant at $\Delta(t=0.9T-T)=\Delta_f$ and the Rabi frequency is changed off from $\Omega(t=0.9T)=\Omega_0$ to $\Omega(t=T)=0$. For the change of $\Omega$ and $\Delta$, the laser pulses are modulated by RF synthesizers (Moglabs XRF) and acousto-optic modulators.
 The whole control operation time is $T=2.5$~$\mu$s, and the optical tweezers are turned off during the operation. The initial and final detuning values are $\Delta_i=-4.0$~$(2\pi)$MHz and $\Delta_f=5.0$~$(2\pi)$MHz. The positions of the atoms we used in the experiment are given in Table~\ref{table_position}. The maximum distance between the coupled atoms that we used is $d=7.6$~$\mu$m, smaller than the Rydberg blockade distance $d_R=(C_6/\Delta_f)^{1/6}=7.7$~$\mu$m for $C_6=1023$~$(2\pi)$GHz$\cdot \mu$m$^6$.

We detect the final states of the Rydberg-atom graph through fluorescence imaging. To achieve this, we reactivate the optical tweezers and initiate the cyclic transition, $\left|5S_{1/2, F=2}\right> \leftrightarrow \left|5P_{3/2, F'=3}\right>$, to capture the photon emissions from all atoms. We employ an electron-multiplying charge-coupled device camera (Andor iXon Ultra 888) for photon detection. In this process, ground-state atoms emit photons, while Rydberg-state atoms are repelled by the optical tweezers and do not emit photons.

\section{Experimental results} \label{Result} 
\begin{figure*}
    \centering
    \includegraphics[width=2.0\columnwidth]{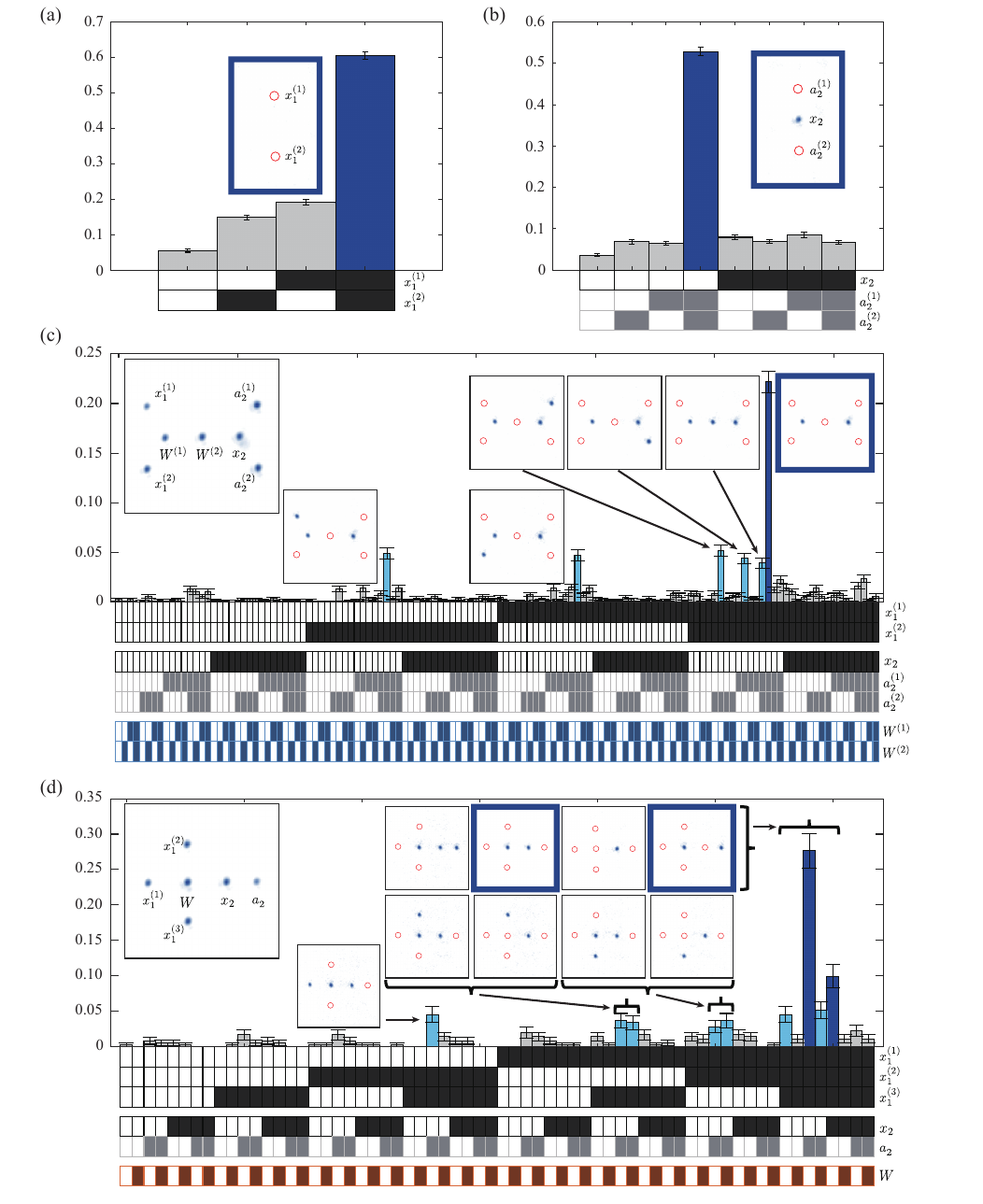}
    \caption{(a) The experimental probability distribution of the data-qubit subgraph $G_1$ for $f_1=-2x_1$. (b) The data-and-offset subgraph $G_2$ for $f_2=x_2$, where $x_2$ and $a_1^{(1,2)}$ are date and offset qubits, respectively. (c) The Rydberg-atom graph $G_3$ for $f_3=f_1+f_2+x_1x_2 = -2x_1+x_2+x_1x_2$, where $W^{(1,2)}$ are the ``even-atom'' quantum-wire. (d) The Rydberg-atom graph $G_4$ for $f_4=f_1+f_2-x_1x_2 = -2x_1+x_2-x_1x_2$, where $x_1^{(1,2,3)}$, $x_2$, $a_2$, and $W$ are the data qubits for $x_1$ and $x_2$, the offset qubit for $x_2$, and the ``odd-atom'' quantum wire, respectively.}
\label{Fig2}
\end{figure*}

\noindent
First, we test the four building blocks of Rydberg-atom graphs for the working principle of Rydberg-atom graph-based QUBO implementation. Figure~\ref{Fig2} shows the experimental results of the graphs $G_1$, $G_2$, $G_3$, and $G_4$ in Figure.~\ref{Fig1}, in which the experimentally observed probabilities are plotted for all possible binary configurations of the Rydberg-atom graphs in their final state (i.e, after the adiabatic quantum control). {The color blocks means qubit state depend on the atoms. Uncolored (colored) blocks means atoms in the ground state $\left|0\right>$ (Rydberg state $\left|1\right>$). } The atom positions of all Rydberg-atom graphs are listed in Table~\ref{table_position}.

\begin{table}[bt]
\centering
\caption{Atom positions of Rydberg-atom graphs}
\begin{ruledtabular}
\begin{tabular}{@{}cclll@{}}
Graphs & Atoms & \multicolumn{3}{l}{Positions $(x,y)$ [$\mu$m]}   \\
\hline\hline

\multirow{1}{*}{$G_1$} &$x_1^{(1,2)}$:& (0.0, $\pm $7.6) \\
\hline

\multirow{1}{*}{$G_2$} & $x_2$:& (0.0, 0.0) & $a_2^{(1,2)}$:&  (0.0, $\pm$7.6)\\

\hline
\multirow{3}{*}{$G_3$} &$x_1^{(1,2)}$:& (-11.4, $\pm $6.6)\\
& $x_2$:& (7.6, 0.0) &$a_2^{(1,2)}$: &  (11.4, $\pm$6.6)\\
& $W^{(1)}$:& (-7.6, 0.0) & $W^{(2)}$:& (0.0, 0.0) \\

\hline
\multirow{3}{*}{$G_4$}  & $x_1^{(1)}$:& (-7.6, 0.0) & $x_1^{(2,3)}$: & (0.0, $\pm$7.6) \\
& $x_2$:& (7.6, 0.0) &$a_2$: & (13.3, 0.0)\\
& $W$:& (0.0, 0.0)  \\

\hline
\multirow{3}{*}{$G'_5$} & $\tilde{W}^{(1)}$:& (-11.4, 0.0) & $\tilde{W}^{(2)}$:& (-3.8, 0.0)\\
 & $x_1^{(1,2)}$:& (-15.2, $\pm$6.6) \\
& $x_2$:& (12.9, 0.0) & $a_2^{(1,2)}$: & (18.3, $\pm$5.4) \\
& $W_1^{(1)}$:& (0.0, 6.6) & $W_1^{(2)}$: & (7.5, 5.4)\\
& $W_2^{(1)}$:& (0.0, -6.6) &$W_2^{(2)}$: & (7.5, -5.4)\\

\hline
\multirow{3}{*}{$G'_6$} & $\tilde{W}^{(1,2)}$:& (-5.4, $\pm$5.4) & $\tilde{W}^{(3)}$: & (0.0, 0.0)\\
& $x_1^{(1,4)}$:& (-5.4, $\pm$13.0) &$x_1^{(2,3)}$: &(-13.0, $\pm$5.4) \\
& $x_2$:& (10.7, 0.0) \\
& $W_1$:& (5.4, 5.4) & $W_2$: & (5.4,  -5.4)\\

\hline
 & $x_1^{(1)}$:& (-18.3, 0.0) & $x_1^{(2)}$:& (-7.6, 0.0) \\
 & $x_2$:& (12.9, 5.4)  \\
{$G_7$} &$x_3$: & (12.9, -5.4) & $a_3$: & (18.3, -10.7) \\
& $W_{1}^{(1)}$:& (-12.9, 5.4) &$W_{1}^{(2)}$: & (-7.6, 10.7) \\ 
& $W_{1}^{(3)}$:& (0.0, 10.7) &$W_{1}^{(4)}$: & (7.6, 10.7) \\
& $W_{2}^{(1)}$:& (-12.9, -5.4) & $W_{2}^{(2)}$:& (-7.6, -10.7) \\
& $W_{2}^{(3)}$:& (0.0, -10.7) & $W_{2}^{(4)}$:&(7.6, -10.7) \\
& $W_{3}^{(1)}$:& (7.6, 0.0) & $W_{3}^{(2)}$:& (18.3, 0.0)  \\

\hline
$G_{\rm LNK}$  & $x_1$:& (-3.8, 3.8) & $x_2$:&(3.8, -3.8) \\
& $W_1$:& (3.8, 3.8) & $W_2$:& (-3.8, -3.8)\\

\hline
$G_{\rm NOT}$  & $x_1$:& (-7.6, 0.0) & $x_2$:&(7.6, 0.0) \\
& $W_1^{(1,2)}$:& ($\pm$3.8, 6.6) &$W_2^{(1,2)}$:& ($\pm$3.8, -6.6)\\
\end{tabular}
\end{ruledtabular}
\label{table_position}
\end{table}

Figure~\ref{Fig2}(a) shows the resulting probability distribution for the Rydberg-atom graph $G_1$, which encodes the function $f_1=-2x_1$.  The atoms are labeled with $x_1^{(1,2)}$ for the $x_1$ data-qubit graph. The colored peaks correspond to $(x_1^{(1)}, x_1^{(2)})=(1,1)$ which result in $x_1=1$, the correct QUBO solution of $f_1$, where the data qubits $x_1^{(1,2)}$ print the same value. In Figure.~\ref{Fig2}(b), the probability distribution is shown for $G_2$, which encodes $f_2=x_2$, where the atoms are labeled with $a_2^{(1,2)}$ (auxiliary) and $x_2$ (data) of the $x_2$ data-and-offset subgraph for $f_2$. The colored peaks correspond to $(x_2, a_2^{(1)}, a_2^{(2)})=(0,1,1)$, which result in $x_2=0$, the minimum QUBO solution of $f_2$, where the offset qubits $a_2^{(1,2)}$ print the opposite value of data qubit $x_2$. 

In Figure.~\ref{Fig2}(c), the experimental result of the Rydberg-atom graph $G_3$ is shown for $f_3=-2x_1+x_2+x_1x_2$, where the atoms $W^{(1,2)}$ are the even-atom quantum wire. There are six colored peaks (from left to right), $(x_1^{(1)}, x_1^{(2)}; x_2, a_2^{(1)}, a_2^{(2)};W^{(1)},W^{(1)})$ = (0,1;0,1,1;0,1), (1,0;0,1,1;0,1), (1,1;0,0,1;0,1), (1,1;0,1,0;0,1), (1,1;0,1,1;0,0), and (1,1;0,1,1;0,1). The highest peak is $(1,1;0,1,1;0,1)$ which corresponds to the QUBO solution $(x_1,x_2)=(1,0)$ of $f_3$. 

In Figure.~\ref{Fig2}(d), the probability distribution is shown for $G_4$, which encodes $f_4=-2x_1+x_2-x_1x_2$. There are nine colored peaks (from left to right), $(x_1^{(1)}, x_1^{(2)}, x_1^{(3)}; x_2, a_2;W)$ = (0,1,1;0,1;0), (1,0,1;0,1;0), (1,0,1;0,1;1), (1,1,0;0,1;0), (1,1,0;0,1;1), (1,1,1;0,0;0), (1,1,1;0,1;0), (1,1,1;0,1;1), and (1,1,1;1,0;0).  By additional term comes from odd-atom wire $-x_1x_2$, $f_4$ has one more solution $(x_1,x_2)=(1,1)$ unlike $f_1+f_2$ and $f_3$. Significant peaks are $(1,1,1;0,1;0)$ and $(1,1,1;1,0;0)$ that correspond to the  the QUBO solutions $(x_1,x_2)=(1,0)$ and $(x_1,x_2)=(1,1)$ of $f_4$.

\begin{figure*}
    \centering
    \includegraphics[width=2.0\columnwidth]{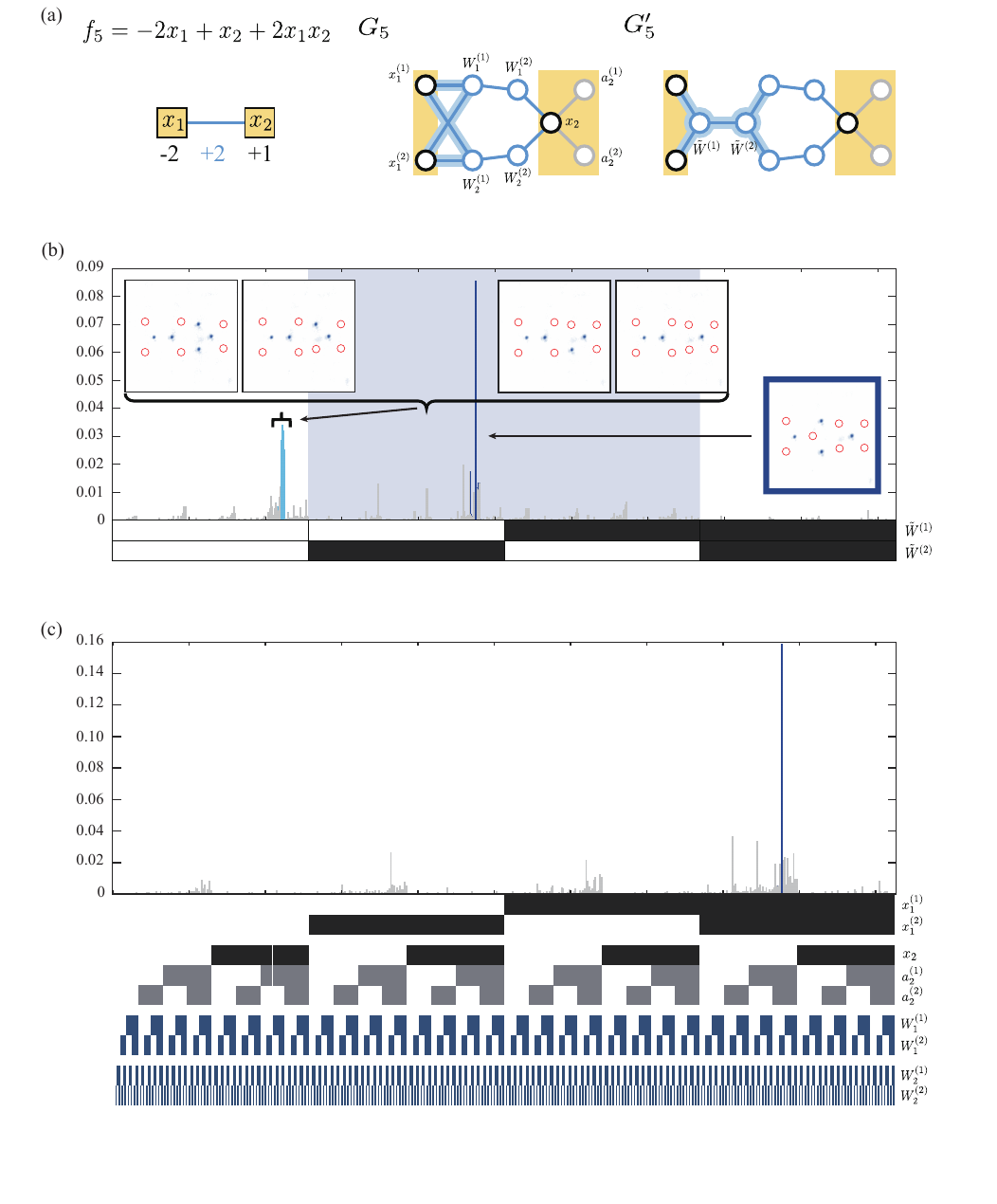}
    \caption{(a) The QUBO function $f_5=-2x_1+x_2+2x_1x_2$, the Rydberg-atom graph $G_5$ (with crossing edges), and $G'_5$ (without crossing edges). (b) The probability distribution of $G'_5$ in all atom states and (c) in the AF-ordered states only. }
    \label{Fig3}
\end{figure*}

\begin{figure*}
    \centering
    \includegraphics[width=2.0\columnwidth]{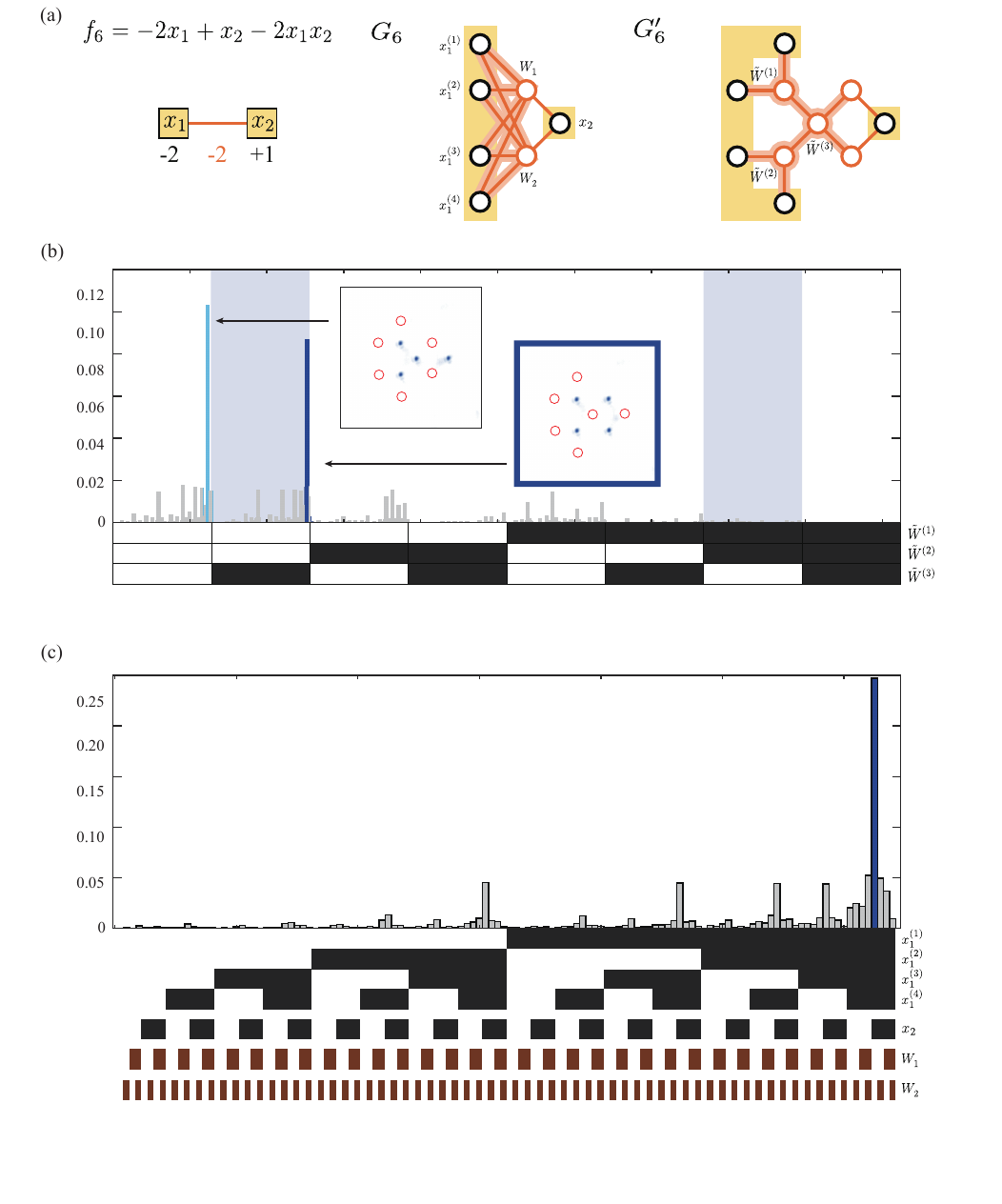}
    \caption{(a) The QUBO function $f_6=-2x_1+x_2-2x_1x_2$, the Rydberg-atom graph $G_6$ (with crossing edges), and $G'_6$ (without crossing edges). (b) The probability distribution of $G'_6$ in all atom states and (c) in the AF-ordered states only.}
    \label{Fig4}
\end{figure*}

Using the examples of $G_1$, $G_2$, $G_3$, and $G_4$ in Figure.~\ref{Fig2}, we have experimentally checked the working principle of the four elementary building blocks of our Rydberg-atom graph QUBO implementation. We now test multiple quantum-wire examples, as shown in Figures.~\ref{Fig3} and \ref{Fig4}, respectively, for $f_5= -2x_1+x_2+2x_1x_2$ and $f_6=-2x_1+x_2-2x_1x_2$. In these demonstrations, we use the compilation method of antiferromagnetic (AF)-ordering quantum-wires, introduced in Ref.~\cite{Byun2022}, to simplify otherwise complicated Rydberg-atom graphs, as $G_5 \rightarrow G'_5$ and $G_6 \rightarrow G'_6$ in Figures.~\ref{Fig3} and \ref{Fig4}, respectively.

Figure~\ref{Fig3}(a) shows an example of two even-atom quantum wires. The Rydberg-atom graph $G_5$ encodes the function $f_5=f_1+f_2+2x_1x_2=-2x_1+x_2+2x_1x_2$. In $G_5$, there is a crossing between the edges $(x_1^{(1)},W_2^{(1)})$ and $(x_1^{(2)},W_1^{(1)})$, which requires three-dimensional arrangement of quantum wires~\cite{Kim2022_wire}. For simplicity, we transform $G_5$ to a two-dimensional $G_5'$, utilizing an AF-ordered quantum wire~\cite{Byun2022}. In order to make the quantum wire $W_{1,2}$ in $G_5'$ to act as the four edges $(x_1^{(1)},W_1^{(1)})$, $(x_1^{(1)},W_2^{(1)})$, $(x_1^{(2)},W_1^{(1)})$, and $(x_1^{(2)},W_2^{(1)})$ in $G_5$, we add two additional vertices $\tilde{W}^{(1,2)}$ and apply the AF-ordering compilation method~\cite{Byun2022}. In Figure.~\ref{Fig3}(b), the full probability distribution of $G'_5$ is shown. The shaded region in Figure.~\ref{Fig3}(b) corresponds to the bit-wise configurations satisfying the AF ordering, i.e., $(\tilde{W}^{(1)},\tilde{W}^{(2)})=(0,1)$ and $(1,0)$ in $G_5'$. There are five significant peaks (from left to right), $(\tilde{W}^{(1)},\tilde{W}^{(2)}; x_1^{(1,2)}; x_2,a_2^{(1,2)}; W_1^{(1)}, W_1^{(2)};  W_2^{(1)}, W_2^{(1)})$ = (0,0;1;0,1;1,0;1,0), (0,0;1;0,1;1,0;1,1), (0,0;1;0,1;1,1;1,0), (0,0;1;0,1;1,1;1,1), and (0,1;1;0,1;0,1;0,1), among which the first four peaks are $(\tilde{W}^{(1)},\tilde{W}^{(2)})=(0,0)$, so they are ignored. In Fig.~\ref{Fig3}(c), the resulting probability distribution is replotted under the AF-ordering condition, i.e., $(\tilde{W}^{(1)},\tilde{W}^{(2)})=(0,1)$ and $(1,0)$, after renormalization. The highest peak in Figure.~\ref{Fig3}(c) is $(x_1^{(1,2)} ; x_2, a_2^{(1,2)}; W_1^{(1)}, W_1^{(2)}; W_2^{(1)}, W_2^{(2)})=(1;0,1;0,1;0,1)$, which results in the QUBO solution $(x_1,x_2)=(1,0)$.

In Figure.~\ref{Fig4}(a), the QUBO function $f_6=-2x_1+x_2-2x_1x_2$ is encoded with the Rydberg-atom graph $G_6$. {The QUBO function could be decomposed as $f_6=f_1+f_2-2x_1x_2=2(x_1+x_2-x_1x_2)+(f_1-2x_1)+(f_2-2x_2)$ by the two odd-atom quantum wire terms $2(x_1+x_2-x_1x_2)$. The data qubit representing $x_1$ increase as $f_1-2x_1=-4x_1$ by four atoms $x_1^{(1,2,3,4)}$. To representing $f_2-2x_2=-x_1$, only one data qubit $x_2$ is needed. Two quantum wires $W_{1,2}$ are used to couple $x_1$ and $x_2$.  Both $W_{1}$ and $W_{2}$ are connected with five vertices $x_1^{(1,2,3,4)}$ and $x_2$ in $G_6$. Implementing high-degree vertices are complex due to the Rydberg blockade. Therefore, the vertex-splitting method introduced in Reference.~\cite{Kim2022_wire} is used to reduce the coupling complexity. We split five-degree vertices $W_1$ and $W_2$ in $G_6$ to two three-degree vertices $\tilde{W}^{(1)}$ and $\tilde{W}^{(2)}$ in $G'_6$ by using quantum wire $(\tilde{W}^{(1)}, \tilde{W}^{(3)})$ and $(\tilde{W}^{(2)}, \tilde{W}^{(3)})$. The vertex splitting method also contains AF-ordering constraints, so that quantum wires follow $(\tilde{W}^{(1,2)}, \tilde{W}^{(3)})=(0,1)$ or $(1,0)$. }
In Figure.~\ref{Fig4}(b), there are two significant peaks, $(\tilde{W}^{(1,2)},\tilde{W}^{(3)}; x_1^{(1,2,3,4)};x_2;W_{1,2})$ = (0,0;1;0;1) and (0,1;1;1;0). The latter one is located in the region satisfying the AF-quantum wire condition (the shaded regions). The renormalized distribution in Figure.~\ref{Fig4}(c) shows that $( x_1^{(1,2,3,4)};x_2;W_{1,2})=(1;1;0)$ is the QUBO solution $(x_1,x_2)=(1,1)$ of $f_6$.

\begin{figure*}
    \centering
    \includegraphics[width=2.0\columnwidth]{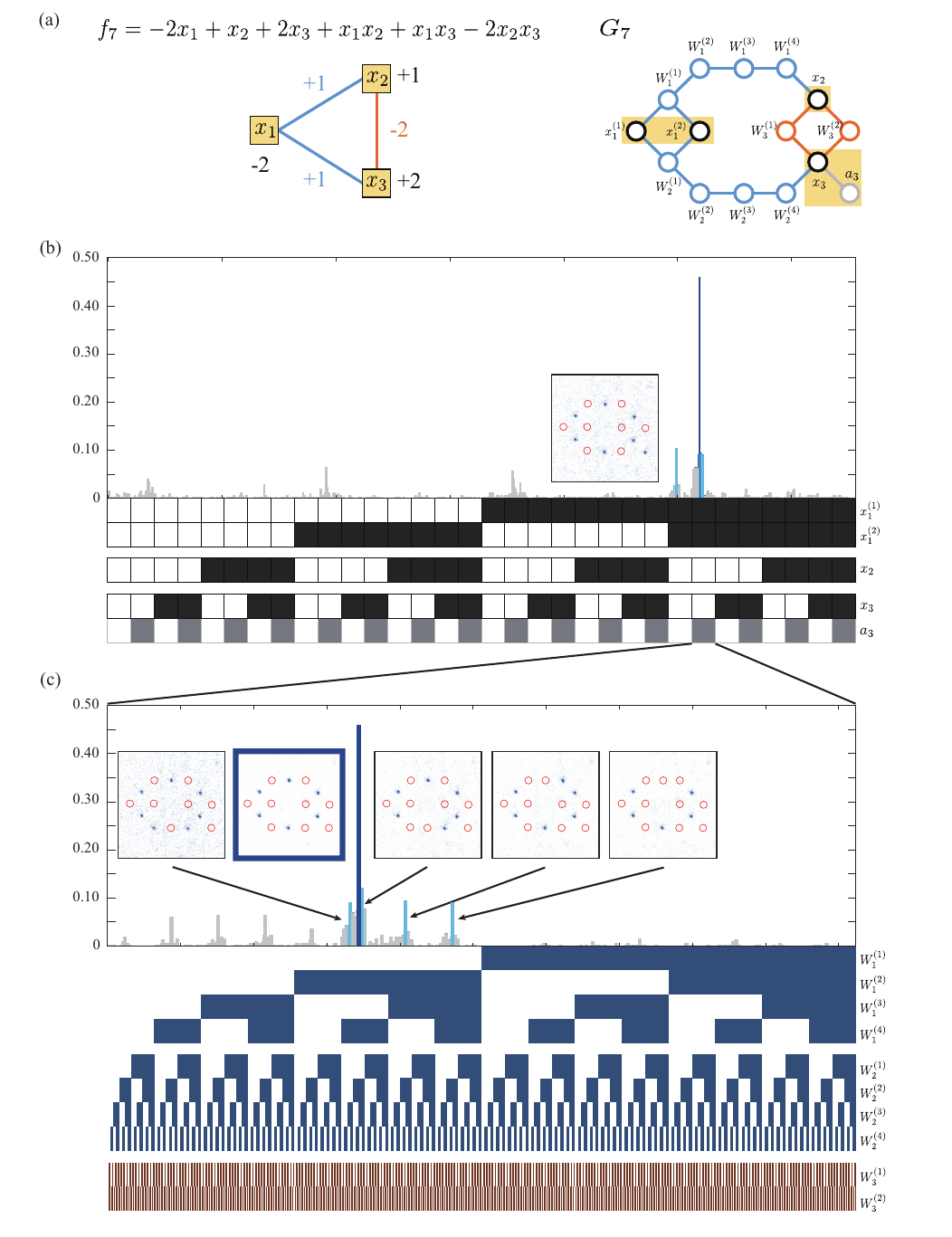}
    \caption{(a) The QUBO function $f_7=-2x_1+x_2 +2x_3 + x_1x_2 + x_1x_3 -2x_2x_3$ and the Rydberg-atom graph $G_7$. (b) The probability distribution of $G_7$ in all atom states and (c) in the $(x_1^{(1)},x_1^{(2)};x_2;x_3,a_3)=(1,1;0;0,1)$ configuration only.
}    \label{Fig5}
\end{figure*}

Furthermore, we test a three-variable QUBO example, 
{$ f_7=-2x_1+x_2+2x_3 +x_1x_2 + x_1x_3 -2x_2x_3=-2x_1+(x_2-2x_2)+(2x_3-2x_3)+x_1x_2+x_1x_3+2x_3+2(x_2+x_3-x_2x_3)$ in Figure.~\ref{Fig5}, which requires a data-qubit subgraphs for $x_1$ and $x_2$, a data-and-offset subgraph for $x_3$, two ``even-atom'' quantum wires $W_1$ and $W_2$, which connect $(x_1, x_2)$ and $(x_1,x_3)$ each, and two ``odd-atom'' quantum wires $W_3^{(1)}$ and $W_3^{(2)}$ for $(x_2,x_3)$.}
The combined Rydberg-atom graph is $G_7$ shown in Figure.~\ref{Fig5}(a). The experimental probability distribution is displaced in Figure.~\ref{Fig5}(b). There are six significant peaks, sharing the states $(x_1^{(1,2)};x_2;x_3)=(1,0,0)$, $(a_3; W_1^{(1)}W_1^{(2)}W_1^{(3)}W_1^{(4)}; W_2^{(1)}W_2^{(2)}W_2^{(3)}W_2^{(4)}; W_3^{(1,2)})$ = (0;0101;0101;1), (1;0101;0010;1), (1;0101;0101;1), (1;0101;0110;1), (1;0110;0101;1), and (1;0111;0101;1). The last five peaks are detailed from the state block $(x_1^{(1,2)};x_2;x_3,a_3)=(1;0;0,1)$, as shown in Figure.\ref{Fig5}(c). The significant peaks are the one or two bit-flip solutions. The highest peak is $(x_1^{(1,2)};x_2;x_3, a_3;W_1; W_2; W_3^{(1,2)} )=(1;0;0,1;0101;0101;1)$, which corresponds to the QUBO solution $(x_1,x_2,x_3)=(1,0,0)$ of $f_7$.

\section{Discussion and conclusion} \label{Discussions} \noindent
Our Rydberg-atom graph implementation of the QUBO problem has focused on encoding the quadratic polynomial function with only atom arrangements and without local addressing of the atoms. Now we turn our attention from the QUBO problem to a graph optimization problem. The QUBO problem in our Rydberg-atom graph approach corresponds to the MIS problem for graphs with weighted vertices and edges. Therefore QUBO could be understood as the maximum weighted independent set (MWIS) problem, which aims to find the set of non-adjacent vertices with the maximum sum of weights on the vertices, of the doubly-weighted graph (vertex-and edge-weighted graph). There are proposals to implement the MWIS problem using locally addressed Rydberg atoms~\cite{Nguyen2022_mwis, Lanthaler2023_mwis, Stastny2023}. 

\begin{figure*}
    \centering
    \includegraphics[width=2.0\columnwidth]{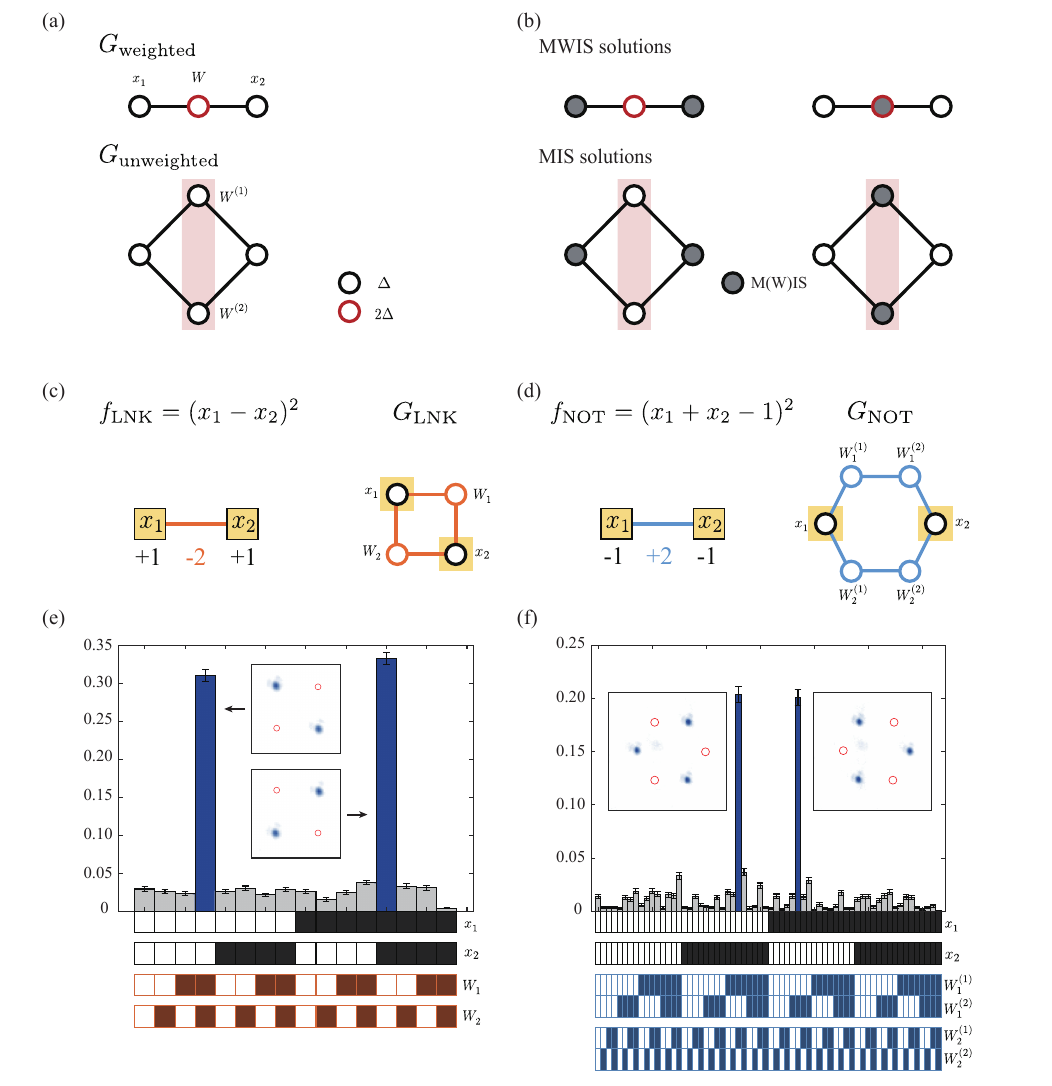}
    \caption{(a) A vertex-weighted graph $G_{\rm weighted}$ is converted to an unweighted graph $G_{\rm unweighted}$ for the maximum weighted independent set (MWIS) problem. (b) The MWIS solutions of $G_{\rm weighted}$ (top) and the MIS solutions of $G_{\rm unweighted}$ (bottom). (c) A QUBO function $f_{\rm LNK}=(x_1-x_2)^2$ for the LINK-constraint $(x_1=x_2)$, the corresponding Rydberg-atom graph $G_{\rm LNK}$.
(d) A QUBO function $f_{\rm NOT}=(x_1+x_2-1)^2$ for the NOT-constraint $(x_1=1-x_2)$, the corresponding Rydberg-atom graph $G_{\rm NOT}$. (e) The experimental probability distribution of $G_{\rm LNK}$ and (f) $G_{\rm NOT}$.
}
    \label{Fig6}
\end{figure*}

We seek a non-addressing version of the Rydberg-atom graph approach to the MWIS problem. As an example of using a Rydberg-atom graph (an unweighted graph) to implement a vertex-weighted graph, we consider the Rydberg-atom Hamiltonian for the MWIS problem given by
\begin{equation}\label{HRyd_MWIS}
\hat{H}_{\rm MWIS}= \sum _{i\in V} \frac{\Omega_i}{2} \hat{\sigma}_{x,i}- \sum _{i\in V} \Delta_i \hat{n}_{i}+\sum _{(i,j)\in E} U_{i,j} \hat{n}_{i} \hat{n}_{j},
\end{equation}
where the laser-frequency detuning $\Delta_i$ defines the vertex-weight of the $i$-th atom (or the vertex). In Figure.~\ref{Fig6}(a), the target graph $G_{\rm weighted}$ has three vertices with the center vertex (denoted by $W$) has a weight ($\Delta_W=2\Delta$) two times bigger than the weights ($\Delta$) of the others ($x_1$ and $x_2$). This weight multiple can be simply implementable using the analogy of the data-qubit subgraph in Section.~\ref{GRAPH}, as shown in Figure.~\ref{Fig6}(a), where the data-qubit atom $W$, which is implemented by two atoms ($W^{(1,2)}$). The resulting many-body ground states of $G_{\rm unweighted}$ are $(x_1,W^{(1)},W^{(2)},x_2)$ = (1,0,0,1) and (0,1,1,0), which correspond to the MIS solutions $(x_1,W,x_2)$ = (1,0,1) and (0,1,0) of the given MWIS problem ($G_{\rm weighted}$), as illustrated in Figure.~\ref{Fig6}(b).

The MWIS solution of a given graph could be interpreted as ``constraints'', the Boolean functions with input and corresponding output bits~\cite{Nguyen2022_mwis, Lanthaler2023_mwis, Stastny2023}. {Our QUBO representation contains MWIS, therefore Boolean constraint gates are implemented by globally driven Rydberg-atom graphs.} 
For example, the LINK-constraint $x_1=x_2$ that delivers the input bit ($x_1$) to the output bit ($x_2$) could be expressed by the QUBO cost function $f_{\rm LNK}=(x_1-x_2)^2$, which corresponds to the Rydberg-atom graph $G_{\rm LNK}$ (see Figure.~\ref{Fig6}.(c)) that results in the correct QUBO solutions, $(x_1,x_2)=(0,0)$ and $(1,1)$ as in Figure~\ref{Fig6}(e). Likewise, as a simple example, the NOT-constraint $x_1=1-x_2$ is represented by the function $f_{\rm NOT}=(x_1+x_2-1)^2$. The corresponding Rydberg-atom graph is $G_{\rm NOT}$ (see Figure.~\ref{Fig6}.(d)) and the QUBO solutions are $(x_1,x_2)=(0,1)$ and $(1,0)$ as in Figure~\ref{Fig6}(f). By examples $G_{\rm LNK}$ and $G_{\rm NOT}$, we check that QUBO's broad representation forms could program constraint satisfaction problems using Rydberg-atom graphs. As applications of QUBO, MWIS problem and constraint satisfaction problem can be programmed via globally driven Rydberg-atom graphs. 

In summary, we have explored the use of Rydberg atoms to tackle the QUBO problem. We have introduced a core set of four essential components: the data qubit, offset qubit, even-atom quantum wire, and odd-atom quantum wire. These building blocks are used to construct Rydberg atom graphs tailored for specific QUBO cost functions. Our proof-of-concept demonstrations have shown the feasibility of employing Rydberg atoms for programming the QUBO problem. The QUBO cost function is adequately represented as a doubly-weighted graph, allowing for a globally controlled Rydberg-atom system operating in the distance-insensitive Rydberg-atom blockade regime to model this weighted graph. Consequently, when these four building blocks are combined to form a Rydberg atom graph, they exhibit the capability not only to address QUBO problems but also to resolve related challenges such as the MWIS problem and the constraint satisfaction problem. Given the well-established and widespread utilization of QUBO as a form of optimization problem, we hope that Rydberg atom graphs are useful to address a multitude of real-world optimization challenges.

The experimental datasets analysed during the current study are available in the figshare repository~\cite{Data}. The authors declare no competing interests.

\end{document}